\documentclass[aps,pre,groupedaddress,superscriptaddress,twocolumn]{revtex4-1}

\usepackage{amsmath}
\usepackage{graphicx}
\usepackage{float}
\usepackage{siunitx}
\usepackage{xcolor}
\definecolor{darkgreen}{rgb}{0,0.6,0.0}

\usepackage{latexsym}
\usepackage{amsmath}
\usepackage{amsfonts}
\usepackage{amssymb}
\usepackage{mathrsfs}
\usepackage{verbatim}
\usepackage{anysize}
\usepackage{soul}

\newcommand{\bo}{\raise-1mm\hbox{\Large$\Box$}}

\begin{document}

\title{Spinning rigid bodies driven by orbital forcing: The role of dry friction}
\normalsize
\author{Pablo de Castro}
\email[]{pdecastro@ing.uchile.cl}

\affiliation{Departamento de F\'isica, FCFM, Universidad de Chile, Santiago, Chile}

\author{Tiago Ara\'ujo Lima}
\email[]{tiago.arj@gmail.com}

\affiliation{Ser Educacional S.A., Recife, Pernambuco, Brazil}
\affiliation{Departamento de F\'{\i}sica, Universidade Federal de Pernambuco, Recife, Pernambuco 50670-901 Brazil}

\author{Fernando Parisio}
\email{fernando.parisio@ufpe.br} 

\affiliation{Departamento de F\'{\i}sica, Universidade Federal de Pernambuco, Recife, Pernambuco 50670-901 Brazil}

\date{\today}

\begin{abstract}

A ``circular orbital forcing'' makes a chosen point on a rigid body follow a circular motion while the body spins freely around that point. We investigate this problem for the planar motion of a body subject to dry friction. We focus on the effect called \emph{reverse rotation} (RR), where spinning and orbital rotations are antiparallel. Similar reverse dynamics include the rotations of Venus and Uranus, journal machinery bearings, tissue production reactors, and chiral active particles. Due to dissipation, RRs are possible only as a transient. Here the transient or \emph{flip} time $t_\textrm{f}$ depends on the circular driving frequency $\omega$, unlike the viscous case previously studied. 
We find $t_\textrm{f}\sim\omega^{\gamma-1}\mu^{-\gamma/2}$, where $\mu$ is the friction coefficient and $\gamma=0$ ($\gamma=2$) for low (high) $\omega$.
Whether RRs really occur depends on the initial conditions as well as on $\mu$ and $H$, a geometrical parameter. The critical $H_\textrm{c}(\mu)$ where RRs become possible follows a $q$-exponential with $q\simeq1.9$, a more restrictive RR scenario than in the wet case. We use animations to visualize the different dynamical regimes that emerge from the highly nonlinear dissipation mechanism of dry friction. Our results are valid across multiple investigated rigid body shapes.

\end{abstract}


\maketitle

\section{Introduction}

Venus rotates around itself in the opposite direction to its rotation around the Sun \cite{correia1,correia2}. The same occurs with Uranus \cite{cameron1975cosmological}. The magnitude and the \emph{sign} of the ratio between the spinning and orbital angular velocities of a planet (or a star) can remarkably alter the tidal instabilities that take place inside its liquid core \cite{le2010tidal}. Dynamical behaviors where the spinning and orbital rotations of a body are antiparallel are dubbed \emph{reverse rotations} (RRs). Examples include bodies inside rotating chambers filled with viscous fluids \cite{Seddon2006,Sun2010,Merlen2011}, the parametrically-excited damped pendulum \cite{Yoshida1998}, the dynamics of bearings of journal machinery \cite{Camillo2006}, chiral active particles \cite{bechinger2016active}, and the problem of biological tissue production. In the latter, 
a common method to generate tissue is the rotating vessel bioreactor, which consists of a container filled with a nutrient-rich medium rotating about its longitudinal axis at constant angular speed. Inside the vessel, a porous disk seeded with cells to be cultured is placed. The rotating fluid keeps the growing tissue suspended against gravity, leading to nontrivial dynamical regimes \cite{Cummings2007}. The existence of a transient between RR and normal spinning regimes can produce ``topological'' defects in the tissue \cite{Cummings2009}. This kind of transient RR can be significantly prolonged if the surrounding fluid is not sufficiently viscous \cite{de2014role}.

Because of their complexity and variants, RRs need first to be studied in simpler settings. For that purpose, a two-dimensional (2D) frictionless model was devised \cite{Parisio2008}. It consists of a 2D rigid body moving on the horizontal plane as a result of \emph{circular driving}. The exerted force is such that a chosen point ($P$) on the body, located \textit{off} the center of mass (CM), acquires a constant angular speed. The body rotates freely around the pivot point $P$ and is subject only to this orbital driving force. RRs are defined as occurring when the CM follows a bounded trajectory in, say, the clockwise direction and, at the same time, the intrinsic (or spinning) angular degree of freedom evolves counterclockwise, or vice-versa. For this minimal model, an intricate analysis of the equation of motion is already required. A separatrix between normal and reverse rotations was found in the parameter space of geometric and initial conditions \cite{Parisio2008}. To extend this problem to more realistic scenarios, friction has to be considered \cite{mizue2012effects}. The case of a rigid body subject to circular driving as well as to viscous forces was worked out in Ref.~\cite{de2014role}. Contrary to what happens in the frictionless case, where steady RRs are possible, this behavior may exist only as a transient when dissipation is considered. The critical value of a geometrical parameter determines where RRs are possible. Against viscosity, this parameter follows a decaying $q$-exponential such that the region where RRs are possible becomes increasingly smaller with friction.

Here we investigate a circularly driven rigid body in 2D under dry (Coulomb) friction. On a rough horizontal surface, the rigid body thus acquires spinning and orbital motions in response to a circular orbital driving. Our motivation is to probe whether and how the \textit{nature} of the dissipation mechanism can suppress or enhance RRs, hence characterizing their transient. In high-load mechanical systems such as certain types of bearings, valves, wheels, and brakes, viscous forces may become negligible, depending on the viscosity and speed regimes at play \cite{olsson1998friction}.  Dry friction may then dominate the dynamics (along with driving forces) as a result of unavoidable microscopial asperities between surfaces in strong contact. As we shall see, a highly nonlinear friction term appears in the angular equation of motion. This is because the dry friction law is independent of (and thus normalized by) the speed of the rigid body, i.e., only the direction of the motion matters, not its magnitude. A richer, non-monotonic behavior emerges for the RR transient period as a function of the circular driving frequency. Moreover, we find a new $q$-exponential behavior for the critical geometrical parameter vs.\ the friction coefficient, whose consequences are discussed below. In the following we provide a full numerical analysis of this dissipation problem and analytical results for the transient RR time. The present work fills an important gap in the rich field of low-dimensional classical mechanics \cite{Farkas2003} as transient RRs are a fundamental problem in rotational terminal dynamics. 

This paper is organized as follows. In Section \ref{System} we present our simplest setting---a rigid rod-mass body with its entire mass concentrated at the CM---and derive its angular equation of motion. In Section \ref{RR} we characterize several aspects of the associated reverse rotations, including where they occur and for how long, both analytically and numerically. In Section \ref{extendedSection} we turn our attention to ellipse-shaped rigid bodies with extended mass distributions and numerically verify our main results from the previous sections in this new, more general setting. This provides strong theoretical evidence that our findings are valid across broad families of rigid bodies. In Section \ref{Conc} we provide our conclusions and a discussion on future directions. The Appendixes give additional details of the derivations and analyses.

\begin{figure}
	\centering
	\includegraphics[width=0.95\columnwidth]{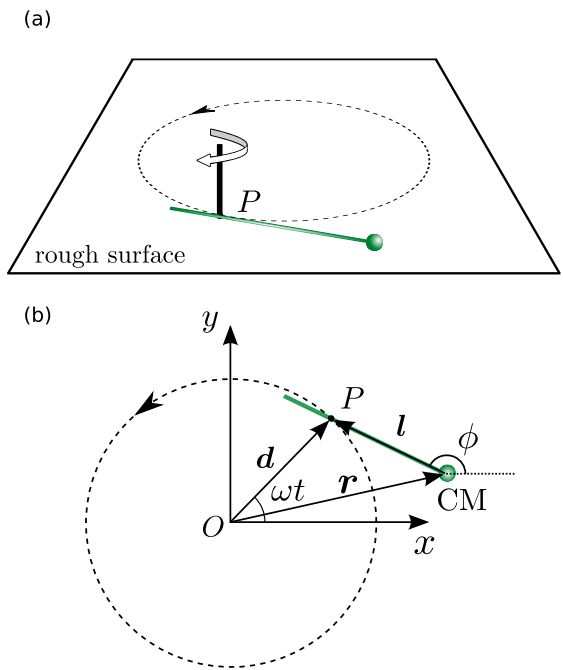}
	\caption{(a) Pictorial representation of the circularly-driven motion of a point mass $m$ attached to a thin, massless rod. The driving is performed by rotating a vertical rod inserted at a point $P$ on the rod-mass system. The mass experiences dry friction due to contact with a rough surface. The rod-mass system rotates freely around the vertical rod at $P$. (b) Geometry of the problem: Schematic upper view of the system, depicting relevant geometric quantities.}
	\label{rod1}
\end{figure}

\section{Rod-mass equation of motion}
\label{System}

Fig.~\ref{rod1}a depicts our  2D rod-mass model. It is composed of a point mass $m$ initially resting on a (horizontal) planar rough surface. Although pointlike, the contact between the mass and the rough surface is assumed to generate finite \textit{dry} friction. That is, the friction force acting on the mass is independent of its speed, depending only on its direction. The mass is attached to a planar rigid massless rod, which together constitute our rod-mass rigid body. Crucially, the system is also subject to
an external planar force, provided by a driving mechanism like a motor. The driving can be thought as exerted through a second thin rod vertically attached to a fixed point $P$ on the body's planar rod. The whole rod-mass system can rotate ``freely'' (except for friction) around $P$. The driving apparatus moves the body by making point $P$ follow a uniform circular trajectory of radius $d$ around a fixed origin ($O$) with angular frequency $\omega$ (see Fig.~\ref{rod1}b). The rigid-body thus acquires an orbital circular motion. Without loss of generality, we assume the orbital rotation to be counterclockwise and use a coordinate system in which point $P$ lies on the positive $x$-axis at $t=0$. We denote the position of the pivot point $P$ by the vector $\boldsymbol{d}$ and the CM (i.e., the point mass) by $\boldsymbol{r}$. Since the rod-mass is assumed to be perfectly rigid, $P$ is always a fixed distance apart from the CM. Their relative position is given by the vector $\boldsymbol{l}$ (of size denoted by $l$), as shown in Fig.~\ref{rod1}b. Finally, the angle between the $x$-axis and the line connecting the CM and $P$ is denoted by $\phi$. The variables $\boldsymbol{r}$ and $\phi$ completely specify the position of the rod-mass system.

In the viscous friction case a circularly-driven rigid \emph{disk} was considered instead of a rod-mass body \cite{de2014role, melo2014efeitos}. For a disk under dry friction, the equation of motion becomes significantly more complicated. The reason is that one has to integrate the dry friction force and torque over the disk's surface. This procedure involves normalizing the velocity since only its direction matters. Thus, intricate square roots arise in the integrand, leading to complicated special functions. A simpler approach which conserves the same physical behavior is to work with the rod-mass system studied here. Yet, extended-mass ellipse-shaped bodies are considered in Section \ref{extendedSection} by solving the integrals numerically at each time instant; they produce indeed the same RR qualitative behavior. Despite the simplicity of the rod-mass system, the problem still needs to be solved mostly numerically. Nonetheless, the derivation of the corresponding equation of motion is much simpler. Furthermore, analytical calculations beyond the equation of motion also become possible.

The dry friction force acting on the mass $m$ is ${\boldsymbol{F}_\textrm{dry}=-\mu m g \hat{v}}$, where $\mu$ is the kinetic friction coefficient between the object and the rough surface, $g$ is the magnitude of the local gravitational acceleration, and $\hat{v}$ is the velocity's direction for the mass. In the inertial frame of reference, Newton's second law reads
\begin{equation}
\boldsymbol{\mathcal{T}_\textrm{total}}\equiv\boldsymbol{d}\times\boldsymbol{F}_\textrm{c}+ \boldsymbol{r}\times \boldsymbol{F}_\textrm{dry}=\frac{\textrm{d}\boldsymbol{L}}{\textrm{d}t},
\label{newton}
\end{equation}
 where $\boldsymbol{F}_\textrm{c}$ is the circular-driving force and $\boldsymbol{L}\equiv m \boldsymbol{r}\times\dot{\boldsymbol{r}}+I_{\textrm{CM}}\dot{\phi}\hat{z}$ is the total angular momentum of the rigid body. The dot denotes time derivative. Since the entire mass of the rigid body is concentrated as a pointlike mass, one has ${I_{\textrm{CM}}\to0}$. The rigid-body constraint ${\boldsymbol{r}+\boldsymbol{l}=\boldsymbol{d}}$ implies that the position of the CM obeys $x(t)=d\cos(\omega t)-l\cos\phi(t)$ and $y(t)=d\sin(\omega t)-l\sin\phi(t)$. Using Newton's second law ${m\ddot{\boldsymbol{r}}=\boldsymbol{F}_\textrm{c}+\boldsymbol{F}_\textrm{dry}}$, the angular equation of motion can be written as (see Appendix \ref{derivation})
\begin{align}
	&\ddot{\phi}-\frac{\omega^2}{H}\sin(\phi-\omega t)\nonumber \\ 
	&-\frac{\mu g}{Hd}\frac{\omega\cos(\phi-\omega t) - H\dot{\phi}}{\sqrt{\omega^2+H^2\dot{\phi}^2-2\omega H \dot{\phi}\cos{(\phi-\omega t)}}}=0,
	\label{motion}
\end{align}
where for convenience we defined the geometric ratio parameter $H \equiv l/d$. In the limit $\mu\to0$ the frictionless case is recovered \cite{Parisio2008}. 

At ${t=0}$ the mass is resting on the horizontal surface and the driving apparatus is switched on. However, the circular-driving force $\boldsymbol{F}_\textrm{c}$ is not compatible with an initial condition where the body is at rest. In order to provide physically valid initial conditions, the driving mechanism is assumed to be robust and the initial dynamics \emph{impulsive}, i.e., point $P$ is taken from rest to the final constant angular velocity much more quickly than any other time scale in the problem. In the frictionless case, it was shown that, given the initial angle $\phi_0$ of the static rigid body, the angular velocity that it acquires immediately after the driving apparatus is turned on is
\begin{equation} 
\label{ic}
\dot{\phi}_0=\frac{\omega}{ H}\cos\phi_0.
\end{equation}
Because of the hypothesis of impulsivity, the fact that friction was not taken into account does not affect the above result: the net force is infinite at $t=0$ regardless of its origin \cite{Parisio2008,de2014role}.

\section{Reverse rotations}
\label{RR}

\begin{figure}[!h]
	\centering
	\includegraphics[width=1\columnwidth]{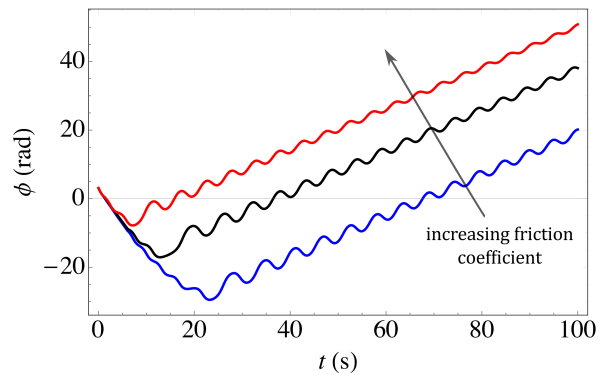}
	\caption{Angle $\phi$ as a function of time $t$ for $\mu g/d=0.033$\,s$^{-2}$, $0.05$\,s$^{-2}$ and $0.1$\,s$^{-2}$ (from bottom to top), $H=0.2$, $\omega=0.6$\,rad\,s$^{-1}$, and $\phi_0=\pi$.}
	\label{phixtpm}
\end{figure}

Fig.~\ref{phixtpm} shows $\phi(t)$ for a rod-mass rigid body as obtained from numerical integration of Eq.~\eqref{motion} with initial condition \eqref{ic}. A decreasing $\phi(t)$ (on average) means that the rigid body is undergoing a RR. Remarkably, despite differences in body shape and friction mechanism, the qualitative dynamical behavior is similar to that of a disk under viscous friction. A transient RRs regime arises before the beginning of a perennial regime of normal rotations. 

\begin{figure*}
	\centering
	\includegraphics[width=\textwidth]{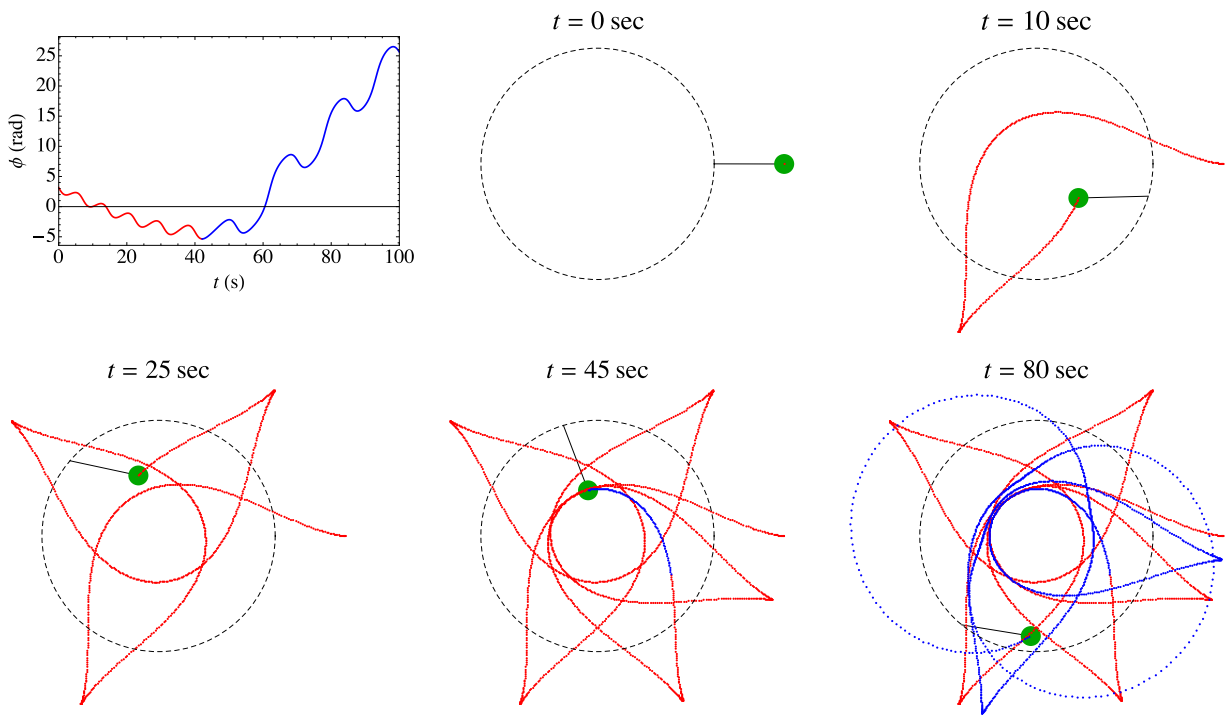}
	\caption{$\phi(t)$ and history of the CM position (green point) for $\mu g/d=0.003$\,s$^{-2}$, $H=0.6$, $\omega=0.6$\,rad\,s$^{-1}$, $\phi_0=\pi$, and times as indicated. For $t<t_\textrm{f}$ ($t>t_\textrm{f}$), $\phi(t)$ and the CM positions are displayed in red (blue).}
	\label{SSs}
\end{figure*}
The RRs last longer for smaller friction coefficients. For an initial condition leading to RR, after a finite time the rigid-body spin invariably flips to a regime of normal rotation. The oscillations in $\phi(t)$ eventually fade out, with $\ddot{\phi}\to0$ as $t\to\infty$. Once this regime is reached, the equation of motion becomes identical to that of a system under infinite friction, for which RRs were formally shown to be absent in the viscous case \cite{de2014role}. To improve our visual understanding of RRs, we show in Fig.~\ref{SSs} (see Movie S1 of the Supplementary Material) the behavior of $\phi(t)$ together with the history of the CM position. For these parameters, the CM motion during RRs follows a series of petal-shaped trajectories characterized by cusps. The existence of these ``petals'' depends on whether the RR-averaged value of $\dot{\phi}$, compared against $\omega$, provides sufficient time for the petals to close up. After the crossover to the normal rotations regime has ended, more rounded trajectories are observed: the internal degree of freedom of the body, $\phi(t)$, synchronizes with the circular forcing as friction has now dissipated sufficient energy. 

\begin{figure}
	\centering
	\includegraphics[width=1\columnwidth]{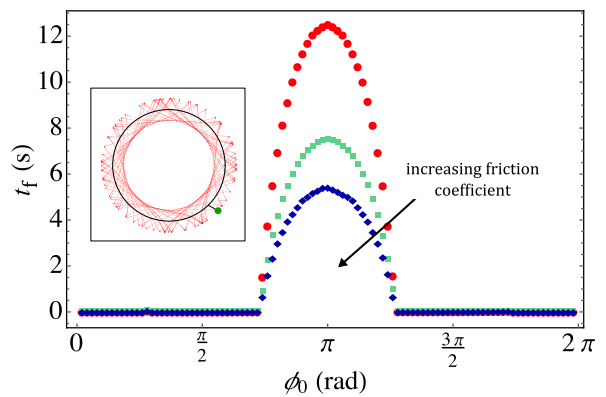}
	\caption{Flip time vs.\ initial condition angle $\phi_0$ for $\mu g/d=7.5$, $12.5$, and $17.5$\,s$^{-2}$, $H=0.25$, and $\omega=100$\,rad\,s$^{-1}$. Inset: Long-time history ($t=250$\,s) of the CM position (green point) undergoing RRs for $\mu g/d=0.0015$\,s$^{-2}$, $H=0.2$, $\omega=0.6$\,rad\,s$^{-1}$, and $\phi_0=\pi$. The black circle represents the uniform circular trajectory of radius $d$ followed by $P$.}
	\label{seco1}
\end{figure}

To characterize the RRs, we identify for which parameters and for how long they occur. The transient or \textit{flip} time $t_\textrm{f}$ that separates the transient RRs from the regime of normal rotations is defined as the global minimum of $\phi(t)$. The initial angle interval that allows for the existence of RRs is insensible to the presence of friction. Neither the magnitude nor the nature of the dissipation mechanism matter \cite{de2014role}. This behavior is displayed in Fig.~\ref{seco1}, where we show $t_\textrm{f}$ vs.\ $\phi_0$ as well as the long-time history of the CM position within the RRs regime. The $\phi_0$ interval is centered at $\pi$, which is the optimal angle for producing long-lived RRs. We will use $\phi_0=\pi$ as our initial condition throughout the rest of our analysis, unless otherwise stated.
In fact, Fig.~\ref{NormalAndPhasePortrait}a (see Movie S2 of the Supplementary Material) shows that normal rotations arise already at $t=0$ when considering the same parameters as in Fig.~\ref{SSs} except with $\phi_0=\pi/2$. Moreover, near the RR transition (which can be crossed by changing either $H$ or $\phi_0$), we find a narrow parameter window where a \textit{transient normal rotation} arises; see Fig.~\ref{NormalAndPhasePortrait}a. To further understand the differences between normal and reverse rotations, the phase portraits for both regimes are shown in Fig.~\ref{NormalAndPhasePortrait}b for the variable $\theta\equiv\phi-\omega t+\pi$, where the two regimes are easily distinguishable.

Let us investigate the behavior of the flip time for low and high $\omega$ as well as its dependence on $\mu g/d$, i.e., the parameter combination that controls the friction term in Eq.~\eqref{motion}. If such relationships are described by power laws, dimensional analysis implies 
\begin{equation}
t_\textrm{f}\sim\omega^{\gamma-1}\left(\mu g/d\right)^{-\gamma/2}
\label{PowerLaw}
\end{equation}
where $\gamma$ is an exponent to be determined. 

\begin{figure}
	\centering
	\includegraphics[width=\columnwidth]{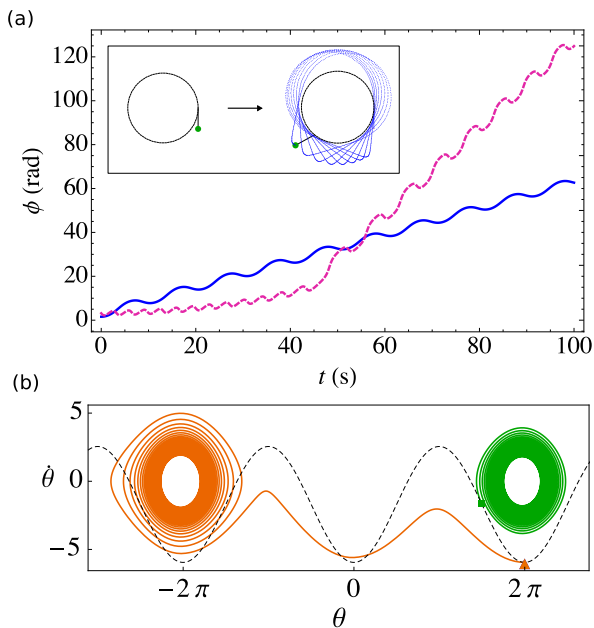}
	\caption{(a) $\phi(t)$ for $\mu g/d=0.003$\,s$^{-2}$, $H=0.6$, $\omega=0.6$\,rad\,s$^{-1}$, and $\phi_0=\pi/2$ (solid blue line) and $\mu g/d=0.003$\,s$^{-2}$, $H=0.8$, $\omega=2.0$\,rad\,s$^{-1}$, and $\phi_0=\pi$ (dashed pink line). Inset: History of the CM position (green point) from ${t=0}$ to ${t=80}$\,s for $\mu g/d=0.003$\,s$^{-2}$, $H=0.6$, $\omega=0.6$\,rad\,s$^{-1}$, and $\phi_0=\pi/2$. The dotted curves represent the uniform circular trajectory of radius $d$ followed by $P$. For these parameters, RRs are not allowed. (b) Phase portrait in the variable $\theta\equiv\phi-\omega t+\pi$ for two initial conditions leading to RRs and normal rotations, $\phi_0=\pi$ (orange triangle) and $\phi_0=\pi/2$ (green square), respectively, for $\mu g/d=0.27$\,s$^{-2}$, $H=0.4$, and $\omega=1.7$\,rad\,s$^{-1}$. The dashed curve is the line of valid initial conditions.}
	\label{NormalAndPhasePortrait}
\end{figure}

We start by considering the limit where $\omega^2 \ll \mu g/d$; see Eq.~\eqref{motion}. The oscillations in $\phi(t)$, which are induced by the orbital forcing frequency $\omega$, have a diverging period in this case. An incomplete RR occurs as the rigid body slowly adjusts $\phi$ towards the beginning of a normal rotation regime, where the external forcing alone will start to completely dictate the dynamics. In that regime, $\phi(t)$ will then increase linearly as $\omega t$ plus a constant. Consequently, the overall shape of $\phi(t)$ is just like that in Fig.~\ref{phixtpm} but without undulations in either the reverse or normal rotation. We therefore approximate $\phi(t)$ in the RR regime by a concave-up parabola constrained to have a minimum at ${t=t_\textrm{f}}$ ($\dot{\phi}=0$) and to the initial conditions, yielding
\begin{equation}
\phi(t) \simeq \phi_0 + \frac{\omega t}{H}\cos(\phi_0) \left ( 1 - \frac{t}{2t_\textrm{f}} \right ),
\end{equation}
which implies $\phi(t_\textrm{f}) \simeq \phi_0 + \omega t_\textrm{f}\cos(\phi_0)/2H$. Inserting into the equation of motion \eqref{motion}, we find
\begin{align}
&\omega\cos(\phi_0) + \omega^2 t_\textrm{f} \sin \left \{ \phi_0 + \omega t_\textrm{f} \left [ \frac{\cos(\phi_0)}{2H} - 1 \right ] \right \} \nonumber \\
&+ (\mu g/d)t_\textrm{f}\cos \left \{ \phi_0 + \omega t_\textrm{f} \left [ \frac{\cos(\phi_0)}{2H} - 1 \right ] \right \} \simeq 0.
\label{PreExpansion}
\end{align}
By solving this equation numerically, we obtained the same qualitative behavior for $t_\textrm{f}$ as from solving Eq.~\eqref{motion} directly without any approximations. Moreover, Eq.~\eqref{PreExpansion} is quantitatively accurate within $10\%$ for a broad range of  numerically calculated flip times. To allow for a closed-form expression for the flip time, we proceed by expanding Eq.~\eqref{PreExpansion} for small $\omega$ with $\phi_0 = \pi$. That gives
\begin{equation}
(\mu g/d)(1+2H)^2 \omega^2 t_\textrm{f}^3 - 8(\mu g/d)H^2 t_\textrm{f} - 8H^2\omega \simeq 0,
\end{equation}
which leads to
\begin{equation}
t_\textrm{f} \simeq \frac{2\sqrt{2}}{\omega(2 + 1/H)} + \mathcal{O}(\omega^1).
\label{SmallOmegaFlipTime}
\end{equation}
To leading order, this expression does not depend on $\mu$, i.e., ${\gamma=0}$ in Eq.~\eqref{PowerLaw}.

In the opposite regime, the case of $\omega^2 \gg \mu g/d$, equation of motion \eqref{motion} becomes frictionless, that is,
\begin{equation}
	\ddot{\phi}-\frac{\omega^2}{H}\sin(\phi-\omega t)=0, 
	\label{highomega}
\end{equation}
in which case RRs are perennial ($t_\textrm{f} \to \infty$) and so $t_\textrm{f}$ increases with $\omega$ for high $\omega$. In this limit, $\gamma$ will be determined from the numerics in two ways, first by plotting $t_\textrm{f}$ vs.\ $\omega$ and then confirmed via $t_\textrm{f}$ vs.\ $\mu g/d$, as follows.

\begin{figure}[!h]
	\centering
	\includegraphics[width=\columnwidth]{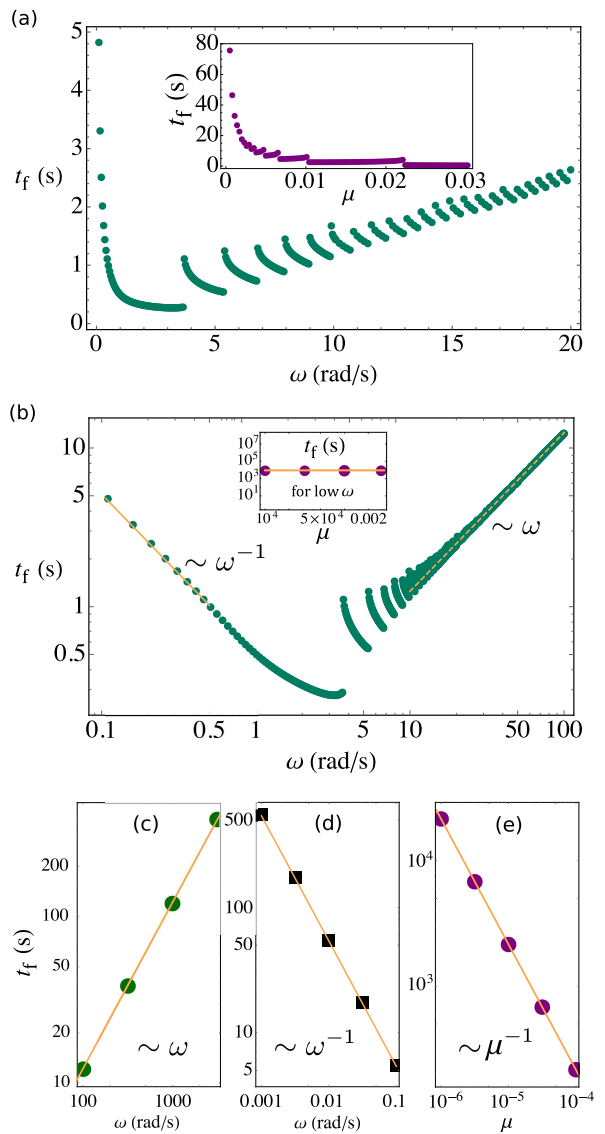}
	\caption{Flip time vs.\ circular driving frequency $\omega$ in (a) linear and (b) log-log scales for $\mu g/d=7.5$\,s$^{-2}$, $H=0.25$, and $\phi_0=\pi$, showing a power law inverse linear behavior for low $\omega$. For higher $\omega$, the flip time increases linearly with $\omega$, indicating the existence of an optimal value of $\omega$ for which the flip time is minimal. The solid lines are power law fits. The inset in (a) shows the flip time vs.\ $\mu$ for $H=0.25$, $g/d=25$\,s$^{-2}$, $\omega=1.0$\,rad\,s$^{-1}$ and $\phi_0=\pi$. (c) Same as (b) but for higher $\omega$ decades. (d) Same as (c) but for low $\omega$. (e) Flip time vs.\ $\mu$ for $H=0.25$, $g/d=25$\,s$^{-2}$, $\omega=0.6$\,rad\,s$^{-1}$ and $\phi_0=\pi$, showing an inverse linear behavior for low $\mu$ (high $\omega$). We also confirmed $t_\textrm {f} \sim \mu^{-1} $ for low $\mu$ and \emph{high} $\omega$ for $H = 0.73$, $g/d = 25$\,s$^{-2}$, $\omega = 200$\,rad\,s$^{-1}$ and $\phi_0=\pi$ in the range of $\mu$ between $0.001$ and $0.03$ (data not shown); for \textit{low} $\omega$, we find numerically that $t_\textrm{f}$ does not depend on $\mu$ [inset in (b)], as predicted by Eq.~\eqref{SmallOmegaFlipTime}.}
	\label{tfxwpm2}
\end{figure}

By numerically solving Eq.~\eqref{motion}, the low-$\omega$ power law behavior in Eq.~\eqref{SmallOmegaFlipTime} is indeed confirmed; see Fig.~\ref{tfxwpm2}d. (The lack of dependence of $t_\textrm{f}$ on $\mu$ for low $\omega$ has also been confirmed through numerics as discussed in Fig.~\ref{tfxwpm2}; data not shown.) The intermediate behavior shows a sequence of discontinuities, which appear when $t_\textrm{f}$ jumps between neighboring local minima. Since those discontinuities are a consequence of our flip time definition, we shall focus on the overall behavior, irrespective of discontinuities. The inset in Fig.~\ref{tfxwpm2}a shows the decreasing behavior of the flip time with $\mu$. Fig.~\ref{tfxwpm2}b highlights the existence of an optimal $\omega$ for which $t_\textrm{f}$ is minimal. This is fundamentally different from the case with viscous friction, where $t_\textrm{f}$ did not depend on $\omega$. The richness in $t_\textrm{f}(\omega)$ steams from the fact that the friction law here does \emph{not} dependent on speed. As a result, for high $\omega$ the energy dissipation rate does not keep up with the increase in $\omega$ and thus the system remains in the RRs regime for increasingly longer periods of time. Finally, Fig.~\ref{tfxwpm2}c shows that for $\omega^2 \gg \mu g/d$ one has $\gamma=2$, as confirmed by Fig.~\ref{tfxwpm2}e.

\begin{figure}[!h]
	\vspace{0.5cm}
	\centering
	\includegraphics[width=\columnwidth]{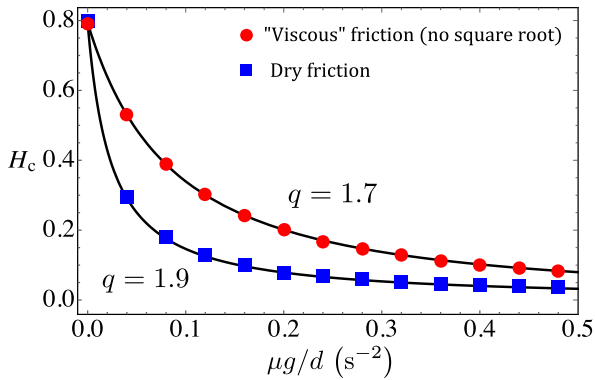}
	\caption{Critical geometrical parameter $H_\textrm{c}$ vs.\ $\mu g/d$. The top curve is calculated using Eq.~\eqref{motion} without the square-root denominator in the friction term, effectively making it viscous. The bottom curve uses Eq.~\eqref{motion}, i.e., the equation of motion for the dry friction case. Both ``viscous'' and dry curves are well adjusted by $q$-exponential functions (solid lines) with $q\simeq1.7$ and $q\simeq1.9$, respectively. Other parameters: $\phi_0=\pi$ and $\omega=0.3$\,rad\,s$^{-1}$.}
	\label{Hc}
\end{figure}
\begin{figure}
	\centering
	\includegraphics[width=\columnwidth]{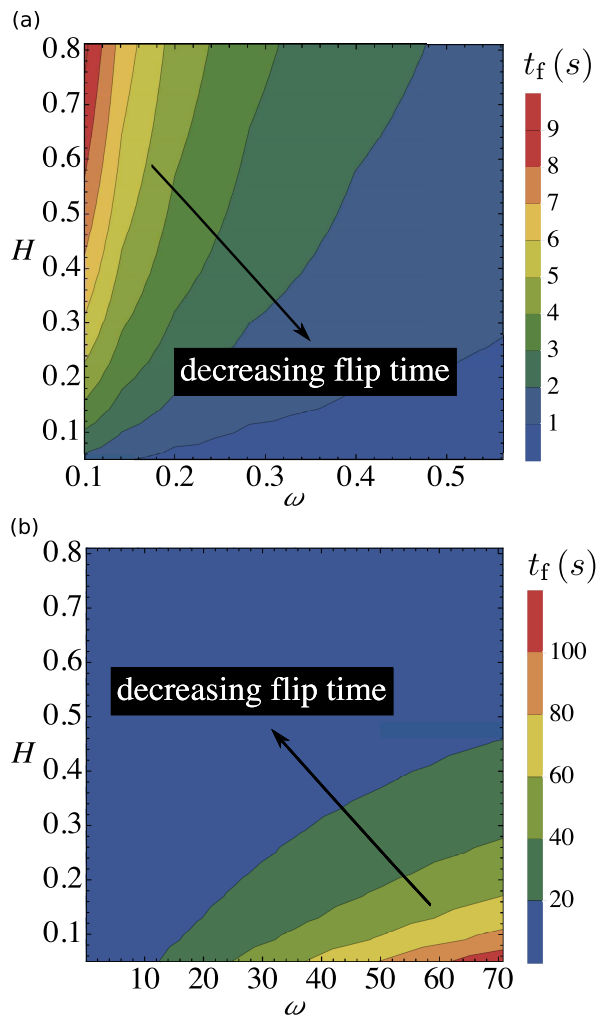}
	\caption{(a) Contour plot of the flip time $t_\textrm{f}$ for $\omega$ vs.\ $H$ with $\mu g/d=1.5$\,s$^{-2}$ for low $\omega$. (b) Same as (a), but including higher values of $\omega$ in order to show the opposite behavior. Notice in particular the opposite behavior for the flip time vs.\ $H$ between low and high $\omega$, as anticipated by Eq.~\eqref{SmallOmegaFlipTime}}
	\label{tfvsHvsomega}
\end{figure}


We now focus on which values of the adimensional geometrical parameter $H=l/d$ allow for RRs. Without dissipation, it was previously found that RRs---which are always perennial in that case---are possible only if $H<H_\textrm{c}(\mu=0)=0.793$ \cite{Parisio2008}. In Fig.~\ref{Hc} we show a geometrical ``phase'' diagram boundary, that is, the critical value of $H$, below which reverse dynamics can occur, as a function of $\mu g/d$. In the viscous case, which is the same as Eq.~\eqref{motion} without the square-root denominator that arises from normalizing the speed, the numerical data follows a $q$-exponential function:
\begin{equation}
\label{qexp0}
	H_\textrm{c}(\mu g/d)=H_\textrm{c}(0)\left[1-\lambda(1-q)\left(\mu g/d\right)\right]^{1/(1-q)},
\end{equation}
with $q=1.724\pm0.003$ and $\lambda=11.64\pm0.01$. In the dry friction case we find that the behavior is similar, albeit with $q=1.936\pm0.004$ and $\lambda=41.32\pm0.09$. Whether there is a deeper explanation for the emergence of $q$-exponentials in these contexts is an open question; one possible simple explanation is that $q$-exponentials can be seen as particularly well suited to fit monotonous functions, \textit{generically} (see Appendix \ref{qexpApp}). In both viscous and dry cases, high friction imposes therefore a condition on $H=l/d$. Such condition is satisfied only for high $d$ (at fixed $l$), showing that friction not only decreases the RRs transient time but also reduces the region in parameter space $d$ vs.\ $l$ where they are possible. In fact, the $q$-exponential for dry friction lies closer to the bottom. In other words, the effect of dry friction is to produce an even more restrictive scenario regarding RRs than in the wet case. This sounds reasonable, \textit{a posteriori}, by noticing that in the viscous case a lubricant fluid surrounding the rigid body is present.

Finally, Fig.~\ref{tfvsHvsomega} shows the flip time as a contour plot for $\omega$ vs.\ $H$. Notice in particular the opposite behavior for the flip time vs.\ $H$ between low and high $\omega$. This had been anticipated by Eq.~\eqref{SmallOmegaFlipTime}. In the scenario where decreasing $\omega$ (or increasing $H$) leads to an increase in the flip time, the RR can also be interpreted as a normal rotation where $\phi(t)$ has not yet reached its first trough (in this case, the picture where RRs require $H<H_\textrm{c}(\mu=0)=0.793$ is no longer valid). One way to think of it is as follows. The time to reach the minimum of $\phi(t)$ in this case increases with $H=l/d$ because, e.g., for high $l$ and fixed $d$ the rigid body takes a long time to complete a spinning cycle around $P$. For high $\omega$, on the other hand, increasing $H$ makes the body spend a longer spinning cycle ``exposed'' to relatively higher dissipation rates, and thus increasing $H$ decreases the flip time.

\section{Extended-mass bodies}
\label{extendedSection}
In this section we turn our attention to rigid bodies with extended (homogeneous) mass distributions rather than systems like the rod-mass body considered above where the entire mass was at the CM. In this case, the equation of motion is altered in a number of ways. First, one has $I_\textrm{CM}>0$. Second, the friction force and torque need to be integrated over the area of the body at each time instant. It is now convenient to define
\begin{equation}
H\equiv \frac{I_\textrm{CM}+m l^2}{m d l}
\end{equation}
and so $\dot{\phi}_0$ can still be written as $\dot{\phi}_0=(\omega/H)\cos{\phi_0}$.

\begin{figure}
	\centering
	\includegraphics[width=0.9\columnwidth]{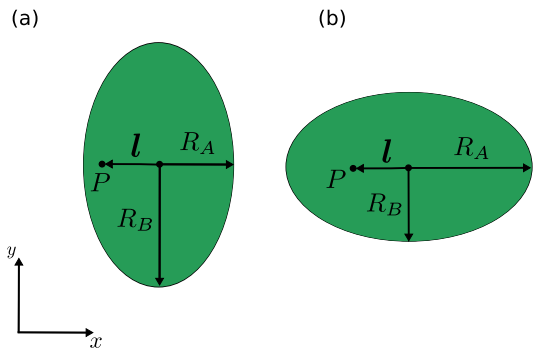}
	\caption{Geometry of the problem for ellipses with semi-axes $R_A=R_0(1+\delta)$ and $R_B=R_0(1-\delta)$ with ${-1<\delta<1}$, showing initial conditions where $\phi_0=\pi$ and the minor axis is along the (a) $x$- and (b) $y$-axis.}
	\label{extended}
\end{figure}

\begin{figure}
	\centering
	\includegraphics[width=\columnwidth]{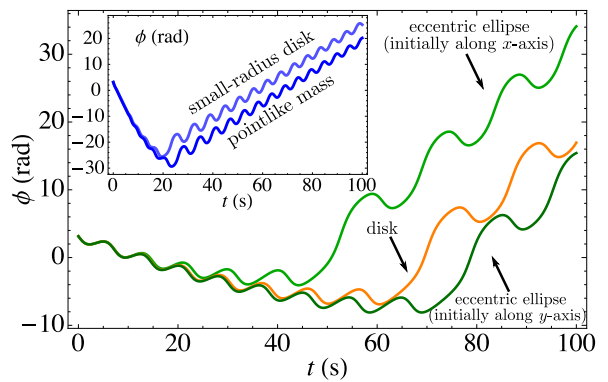}
	\caption{$\phi(t)$ for a disk of radius ${R_0=0.4036}$\,m (orange curve) and two highly eccentric ellipses with ${R_0=0.3}$\,m for $\delta=0.9$ (lighter green curve) and $\delta=-0.9$ (darker green curve). The other parameters are the same for the three curves: $\mu g/d=0.0033$\,s$^{-2}$, $\omega=0.6$\,rad\,s$^{-1}$, ${H=0.607}$, ${l=0.2}$\,m, and ${\phi_0=\pi}$. The mass $m$ is not relevant as it gets canceled out. The inset shows $\phi(t)$ for a small disk with ${R_0=0.05}$\,m and for a pointlike mass, where $\mu g/d=0.033$\,s$^{-2}$, $H=0.2$, ${\omega=0.6}$\,rad\,s$^{-1}$, and $\phi_0=\pi$.}
	\label{ExtendedPhi}
\end{figure}

We consider a family of ellipse-shaped rigid bodies, which of course includes disks. Their moment of inertia and total area are, respectively, ${I_\textrm{CM} = m \left(R_A^2 + R_B^2\right)/4}$ and ${A_{\textrm{tot}}=\pi R_A R_B}$. To investigate deviations from a disk, we consider ${R_A=R_0(1+\delta)}$ and ${R_B=R_0(1-\delta)}$ with ${-1<\delta<1}$, as shown in Fig.~\ref{extended}. Our initial conditions are such that at $t=0$ the semi-axis with size $R_A$ is parallel to the $x$-axis and, as before, we take $\phi_0=\pi$.
The dry friction surface integrals read
\begin{equation}
\boldsymbol{F}_\textrm{dry}=-\frac{\mu m g}{A_\textrm{tot}}\int_{S} \frac{\dot{\boldsymbol{s}}}{|\dot{\boldsymbol{s}}|}\,\textrm{d}S,
\end{equation} 
and
\begin{equation}
\boldsymbol{\mathcal{T}}_\textrm{dry}=-\frac{\mu m g}{A_\textrm{tot}}\int_{S}\boldsymbol{s}\times \frac{\dot{\boldsymbol{s}}}{|\dot{\boldsymbol{s}}|}\,\textrm{d}S,
\end{equation}
where $\boldsymbol{s}$ is a vector from the origin to the position of a mass element in the rigid body and $S$ denotes the surface of the body. Although these integrals are highly complicated to evaluate analytically due to the square roots in the normalized speeds, one can proceed by integrating them numerically. Because the surface integral needs to be performed at each time instant, this numerical procedure is computationally expensive. Nonetheless, we have still calculated $\phi(t)$ and $t_\textrm{f}$ for several values of $\delta$ and the other parameters such that all regions of the parameter space have been properly sampled. The numerical implementation was validated by comparing the dynamics of the rod-mass body for the pointlike mass versus a small disk. In the inset of Fig.~\ref{ExtendedPhi} this is shown for a not-too-small disk, in order to make the two curves distinguishable. 

For $\delta\neq0$, we looked at two perpendicular initial conditions for the orientation of the ellipse on the plane, one of which has the major axis of the body parallel to the $x$-axis of our coordinate system (Fig.~\ref{extended}). The explored parameters also included situations where $P$ is located outside of the body, which is possible through a rigid massless rod connecting them as before. For the three cases illustrated in Fig.~\ref{ExtendedPhi}b, which consist of two highly eccentric ellipses (one for each initial condition in Fig.~\ref{extended}) and a disk, point $P$ is inside the massful part of the body and thus no connecting rod is present. They all have the same parameter values (including $I_{\textrm{CM}}$, $H$, and $l$), except for $R_A$ and $R_B$ (or $R_0$ and $\delta$). Their areas are therefore different: each of the two ellipses have roughly $10$ times less area than the disk.

Despite all changes, we find that the qualitative behavior of $\phi(t)$ (Fig.~\ref{ExtendedPhi}) as well as the values of the power law exponents remain exactly unchanged across all rigid bodies shapes investigated in the present work. For the sake of brevity, any additional data will therefore be omitted here. The fact that the RR dynamical behavior does not depend on the rigid body shape is remarkable since one could expect at first glance to find functional form changes in the integrals leading to new behavior. It could be interesting to investigate whether other shapes lead to distinct qualitative regimes, although our results point in the opposite direction.

\section{Conclusions and discussion}
\label{Conc}

Here we characterized the role of dry (Coulomb) friction in the phenomenon of reverse rotations, where the spinning and orbital angular velocities of a rigid body are antiparallel. We first investigated the simplest possible setting: a circular driving force moving a pivot point $P$ of a rod-mass system that is free to rotate around $P$. This allowed for both numerical and analytical results. We then verified the numerical analysis by generalizing to extended-mass rigid bodies, namely ellipse-shaped bodies and sampled the range of geometric parameters. The transient time $t_\textrm{f}$ of the reverse rotation regime is not independent of the circular driving $\omega$ as in the viscous friction case \cite{de2014role}: in fact, we found analytically and numerically that $t_\textrm{f}\sim\omega^{\gamma-1}\mu^{-\gamma/2}$, where $\mu$ is the friction coefficient and $\gamma=0$ ($\gamma=2$) for low (high) $\omega$. The critical value of the geometrical parameter $H$ ($\equiv l/d$ in the rod-mass system) that defines where reverse rotations are possible---where $l$ is the distance between the driving force application point and the body's CM and $d$ is the driving apparatus arm size---decays as a $q$-exponential with $q\approx1.9$. This is more restrictive with respect to reverse rotations than the viscous friction case. The general reverse rotation behavior found here is independent of rigid body shape for all studied cases.

In future work, it will be interesting to consider gravitational effects, i.e., a tilted surface. For a rod-mass body, this turns the system into a double pendulum \cite{Baker2006} where the inner particle is circularly forced. Our preliminary analyzes of the Poincar\'{e} maps and Lyapunov exponents show that, as for a conventional double pendulum, the system may enter nontrivial dynamical behaviors such as chaos or mixed phase spaces \cite{ott2002chaos, lima2013ergodicity, lima2015classical}. Normal or reverse rotations are also possible.

Another future avenue is the spinning of \textit{chiral active rigid bodies} \cite{bechinger2016active}. Active particles are objects capable of self-propulsion \cite{andrea2020}. In many cases, they move circularly, i.e., with a rotational swimming velocity \cite{bechinger2016active}. One example are rigid L-shaped microparticles coated by a reactive paint, where asymmetrical chemical reactions with a surrounding solvent produce a net active centripetal force \textit{off} the CM \cite{kummel2013circular}. Depending on the details of the self-propulsion mechanism, such \textit{chiral} active particles may be similar to our present system provided that diffusive-like forces are incorporated. The collective behavior of RRs in active matter could be remarkable. On each collision, a new RR could start. For chiral mixtures of bodies with distinct attributes \cite{levis2019simultaneous}, one expects a yet much richer behavior induced by ``polydispersity'', similarly to what occurs with the phase behavior of multicomponent fluids \cite{PabloPeter1, PabloPeter2, PabloPeter3, de2020active}. A fundamental setting would be a binary mixture of bodies that can undergo RRs with others that cannot.

\section*{Acknowledgments}
This research is supported by the Millennium Nucleus Physics of Active Matter of ANID (Chile) and by CNPq, CAPES and FACEPE (Brazil). We thank Paulo C.~Godolphim for a critical reading of the manuscript.

\appendix
\setcounter{equation}{0}
\renewcommand\theequation{A.\arabic{equation}}
\section{DERIVATION OF THE ROD-MASS EQUATION OF MOTION}
\label{derivation}

Here we provide additional details for the derivation of equation of motion \eqref{motion} for the rod-mass system under circular driving and dry friction. As discussed in Section \ref{System}, Newton's second law for rotation is
\begin{equation}
\boldsymbol{d}\times\boldsymbol{F}_\textrm{c}+ \boldsymbol{r}\times \boldsymbol{F}_\textrm{dry}=\frac{\textrm{d}\boldsymbol{L}}{\textrm{d}t},
\end{equation} 
where we can use Newton's second law for translation $m\ddot{\boldsymbol{r}}=\boldsymbol{F}_\textrm{c}+\boldsymbol{F}_\textrm{dry}$ to eliminate $\boldsymbol{F}_\textrm{c}$.
Because of the rigid-body constraint $\boldsymbol{r}+\boldsymbol{l}=\boldsymbol{d}$, where ${\boldsymbol{r}=x\,\hat{x}+y\,\hat{y}}$ and ${\boldsymbol{l}=l(\cos\phi\,\hat{x}+\sin\phi\,\hat{y})}$, we have ${x=d\cos(\omega t)-l\cos\phi}$ and ${y=d\sin(\omega t)-l\sin\phi}$. As the whole mass of the rigid body is pointlike, its moment of inertia with respect to a vertical axis of rotation passing through the CM is ${I_\textrm{CM}=0}$. (With respect to point $P$, the parallel axis theorem gives ${I_P=ml^2+I_\textrm{CM}=ml^2}$.) Since thus $\boldsymbol{L}=m\boldsymbol{r}\times\dot{\boldsymbol{r}}$, we have $\dot{\boldsymbol{L}}=m\dot{\boldsymbol{r}}\times\dot{\boldsymbol{r}}+m\boldsymbol{r}\times\ddot{\boldsymbol{r}}=m\boldsymbol{r}\times\ddot{\boldsymbol{r}}.$ We can now write the dry friction force as
\begin{align}
\boldsymbol{F}_\textrm{dry}&=-\mu m g \frac{\dot{\boldsymbol{r}}}{|\dot{\boldsymbol{r}}|}\nonumber\\
&=\frac{\mu m g}{\sqrt{d^2 \omega ^2+l^2 \dot{\phi}^2-2 d l \omega  \dot{\phi} \cos ( \phi-\omega t )}}\times\nonumber\\
&\left[\left(d \omega  \sin (\omega t)-l \dot{\phi} \sin \phi\right)\,\hat{x}\nonumber\right.\\
&\left.-\left(d \omega  \cos (\omega t)-l \dot{\phi} \cos \phi\right)\,\hat{y}\right]\nonumber\\
\end{align} 
Proceeding similarly for $\ddot{\boldsymbol{r}}$ and inserting everything into the rotational second law, we obtain Eq.~\eqref{motion}, where the definition $H\equiv l/d$ eliminates $l$.

\setcounter{equation}{0}
\renewcommand\theequation{B.\arabic{equation}}
\section{AN ALTERNATIVE LOOK AT {\large $q$}-EXPONENTIALS}
\label{qexpApp}
In this appendix we provide a derivation of $q$-exponentials as objects which are suited to fit monotonous functions.

Suppose one intends to approximate some monotonous function, knowing its value $f(0) \ne 0$ and its derivative $f'(0)$ at $x=0$.
For definiteness, we consider decreasing functions, as those depicted in Fig.~(\ref{Hc}).
In this poor-information scenario the best thing one can do to extrapolate the values assumed by $f(x)$ for $x \ne 0$ is to write 
\begin{equation}
\label{taylor}
f(x) \approx f(0)+f'(0)x. 
\end{equation}
A straight line is a rough approximation to an arbitrary monotonous function. As an extra ingredient, consider that some other piece of information is provided, e.g., the value $f$
assumes for some other point $x \ne 0$ or $f''(0)$, its second derivative at $x=0$. 

Going to the second order term in the Maclaurin series would leave us with a parabola, which is not monotonous. Let us take, instead, a seemingly circular approach, namely, to estimate (also to first order) the function $F(x)\equiv[f(x)]^Q = \tilde{F}(x)+O(x^2)$, and then to write $f(x) \approx [\tilde{F}(x)]^{1/Q} \equiv \tilde{f}(x)$. First we have
\begin{equation}
\tilde{F}(x)= F(0)+F'(0)x, 
\end{equation}
with $F(0)=[f(0)]^Q$ and $F'(0)=Q[f(0)]^{Q-1}f'(0)$. That is,
\begin{equation}
\tilde{F}(x)= [f(0)]^Q+Q[f(0)]^{Q-1}f'(0)x, 
\end{equation}
which amounts to
\begin{equation}
\tilde{f}(x)/f(0)=\left[1-Q\frac{|f'(0)|}{f(0)}x\right]^{1/Q},
\end{equation}
corresponding to our approximation of the actual function $f(x)$.
The above expression is exactly the definition of the $q$-exponential:
\begin{equation}
\label{qexp}
\tilde{f}(x)=f(0)\left[1-Q\lambda x\right]^{1/Q} = f(0) \, {\rm exp}_Q(-\lambda x), 
\end{equation}
with $\lambda=|f'(0)|/f(0)$. Note that we only assumed that $f$ is a steady decreasing function of $x$.
These observations would be mere curiosity if it were not the fact that by expanding the previous expression to first order in $x$ we get $\tilde{f}(x)=f(0) \, {\rm exp}_Q(-\lambda x)\approx f(0)+f'(0)x$, {\it no matter the value of} $Q$. Therefore, we obtain a first order approximation at least as good as (\ref{taylor}), with the difference that there is an adjustable parameter that can be used to improve the extrapolation, if one is given any extra information on $f(x)$, while keeping the monotonous character of the original function.

\bibliographystyle{unsrt}

\bibliography{biblio.bib}

\end{document}